# Resolution limits of extrinsic Fabry-Perot interferometric displacement sensors utilizing wavelength scanning interrogation

Nikolai Ushakov,[1],* Leonid Liokumovich,[1]

[1] *Department of Radiophysics, St. Petersburg State Polytechnical University, Polytechnicheskaya, 29, 195251, St. Petersburg, Russia*
*Corresponding author: n.ushakoff@spbstu.ru

The factors limiting the resolution of displacement sensor based on extrinsic Fabry-Perot interferometer were studied. An analytical model giving the dependency of EFPI resolution on the parameters of an optical setup and a sensor interrogator was developed. The proposed model enables one to either estimate the limit of possible resolution achievable with a given setup, or to derive the requirements for optical elements and/or a sensor interrogator necessary for attaining the desired sensor resolution. An experiment supporting the analytical derivations was performed, demonstrating a large dynamic measurement range (with cavity length from tens microns to 5 mm), a high baseline resolution (from 14 pm) and a good agreement with the model. © 2014 Optical Society of America

## 1. Introduction

Fiber-optic sensors based on the fiber extrinsic Fabry-Perot interferometers (EFPI) [1] are drawing an increasing attention from a great amount of academic research groups and several companies, which have already started commercial distribution of such devices. Sensors demonstrating high sensitivity to a great diversity of measurands [2-6], able to measure combinations of different quantities with minimal cross-sensitivity have been designed. High measurement resolution of such systems is stimulated by the recent progress in demodulation technique [4,7,8], with the best known cavity length resolution of about 30 pm and the measurement range from tens microns to 3-4 mm.

A great diversity of EFPI baseline demodulation techniques has been developed, among them two general classes can be separated: tracking the cavity length variations and capturing the absolute value. The main problem of the approaches of the first class is obtaining the linear response in the interference signal, for that multi-wavelength operation producing quadrature signals [9] and Q-point stabilization techniques [10] are utilized. In the second class the most promising approaches are based on the use of an etalon reference (readout) interferometer [11] and baseline estimation from the registered EFPI spectral function [2,4-8]. With the use of the last approach the best known resolution and dynamic range was reported, also an ability of tracking a system of multiplexed sensors was demonstrated [2]. Among the spectral measurement approaches the frequency scanning interferometry [12,13] is one of the most advantageous.

Despite the great progress in fabrication and applications of different sensors based on EFPI [2,4,14], the analytical study of fundamental limitations on the performance of such sensors isn't well developed and only few publications concerning this question are known [15-17]. However, for the purposes of practical implementation of EFPI-based sensors an analytical description of their resolution limits is of a great interest. Considering the wavelength scanning interrogation, an analysis of errors provoked by instabilities of the laser frequency tuning is present in [18], still only slow and large instabilities are considered.

In this Paper, a mathematical model considering the main limitations on the resolution of EFPI displacement sensor, interrogated by spectrum measurement is developed. The conclusions of the analytical analysis are compared to the results of the experimental study of EFPI baseline resolution.

## 2. Theoretical analysis

Strictly, spectral transfer function of the Fabry-Perot interferometer is defined by the Airy function. For the considered low-finesse interferometers, the dependency can be simplified (by taking into account only two beams) and written in the following form

$$S_{FP}(L, \lambda) = S_0(L, \lambda) + S(L, \lambda), \quad (1)$$

$$S(L, \lambda) = S_m \cos(4\pi n L/\lambda + \gamma(L, \lambda)), \quad (2)$$

where $S_m = 2(R_1 R_2^*)^{1/2}$ – doubled geometrical mean of effective mirrors reflectivities $R_1$ and $R_2^* = R_2 \cdot \eta$, where $\eta$ accounts for the optical losses caused by the light divergence inside the cavity, $n$ is the refractive index of the media between the mirrors, $\lambda$ is the light wavelength. Registered spectral function $S'_i$ ideally equals $S(L_0) = S(L_0, \lambda_i)$, $L_0$ – actual cavity length, $\lambda_i = \lambda_0 + \Delta \cdot i$, where $\lambda_0$ is the central wavelength, $\Delta$ is step between the spectral points, $i = -(M–1)/2...(M–1)/2$, – spectral point number, $M$ is the number of points in the registered digitized spectrum (must be odd for this indexing). In the spectrum analyzer utilized in the current study as well as in all the performed simulations, $M$=20001. For even number of spectrum points this indexing can be easily rewritten. The additional phase term $\gamma(L, \lambda)$ is a superposition of a phase shift induced by the diffraction-induced wavefront deformation (analyzed in Appendix A) and an additional phase term induced by the mirrors.

### 2.1. Initial propositions and task statement

Approaches [15,17] apply modifications of classical phase- and frequency-estimation algorithms to the obtained EFPI spectral function $S'_i$. A representative variant of these algorithms is proposed in [19], utilizing approximation of the measured EFPI spectral function $S'(\lambda)$ by the expression (2) by means of the least-square fitting, minimizing the residual norm

$$R(L) = \|S'(\lambda) - S(L,\lambda)\| = \sqrt{\sum_i [S'_i - S_i(L)]^2} \ . \quad (3)$$

In sequel, the resultant calculated $L$ value providing the global minimum of $R(L)$ function will be denoted as $L_R$. In an ideal case $S'_i = S_i(L_0)$ it's obvious that $L_R$ equals $L_0$. In practice, fluctuations in spectral measurement process cause a disagreement of $S'_i$ and $S_i(L_0)$, resulting in fluctuations of $L_R$ value and hence, limited resolution of the cavity length measurement.

The final baseline measurement resolution will depend on the stability of the utilized spectrometer, rigidity of the EFPI sensing element, and the robustness of the approximation algorithm. Comparing to [19], the modifications made in [7,8] resulted in a great improvement of the baseline measurement resolution. Throughout this Paper signal processing algorithm [8] was used for the baseline calculation.

As the frequency-scanning interrogation is considered, the following parameters of the optical spectrum analyzer limiting EFPI displacement sensor resolution are considered in the developed model:

1. Absolute wavelength scale shift $\Delta\lambda_0$, determined by fluctuations of the triggering of the scanning start, $\sigma_{\Delta\lambda}$=stdev$\{\Delta\lambda_0\}$.
2. Fluctuation of the wavelength scale factor $\delta$, defined by the scanning speed drift.
3. Jitter of the wavelength points $\delta\lambda_i$, caused by the fluctuations of the signal sampling moments, $\sigma_{\delta\lambda}$=stdev$\{\delta\lambda\}$.
4. Additive noises $\delta s_i$, produced by the photo registering units, by the light source intensity noises, etc. $\sigma_s$=stdev$\{\delta s\}$.

On this basis, the measured spectrum will be determined as $S'_i = S(L_0, \lambda_i + \Delta\lambda_0 + \Delta\cdot\delta\cdot i + \delta\lambda_i) + \delta s_i$, i.e. will be distorted comparing to the ideal spectrum $S_i(L_0)$.

The first two factors ("scaling") vary from spectrum to spectrum and their influence on the EFPI displacement sensor resolution can be considered directly. The influence of the 3-rd and the 4-th ("noisy") factors is more complex and produces a distortion of the registered spectrum $S'_i$, hence, the produced $L_R$ deviations will depend on the noise-immunity of the baseline detection algorithm. Therefore, the robustness of the algorithm [8] and the relation between the "noisy" factors and the SNR of the registered spectrum $S'_i$ were studied independently, after that influences of all mechanisms were combined.

It should be noted that for the typical cavity length values $L_0 > 30$ μm and the narrowness of the registered spectral range compared to the central wavelength $\lambda_0$, phase term $\gamma$ can be considered constant in a neighbourhood of $\lambda_0$, resulting in a simplification of further developments with no considerable drawback of the final expressions. In the Appendix B we support the above assumption by analytical and numeric calculations.

### 2.2. Noise sources analysis

Considering the scale shift, one can express $S'_i$ as $S(L_0, \lambda_i + \Delta\lambda_0)$, expand the argument of (2) by powers of $\lambda$, hence, obtaining $4\pi nL/(\lambda_i + \Delta\lambda_0) \approx 4\pi nL/\lambda_i - 4\pi nL\Delta\lambda_0/\lambda_i^2$, resulting in a measurement error

$$\delta L \approx -\Delta\lambda_0 \cdot L_0/\lambda_0. \quad (4)$$

The influence of the scale factor can be considered analogously, assuming $S'_i = S(L_0, \lambda_i + \delta\cdot\Delta\cdot i)$ and taking into account the smallness of $\delta$, $S'_i$ can be expressed as

$$S'_i \cong S_i(L_0) - S_m \frac{4\pi nL_0 \delta \cdot \Delta \cdot i}{\lambda_0^2} \sin\left(\frac{4\pi nL_0}{\lambda_i} + \gamma\right). \quad (5)$$

For relatively small $\delta$, the second term in (5) will only provoke a uniform ascent of $R(L)$ function, with nearly no effect on the fitting result. If an assumption $(\lambda_i - \lambda_0)/\lambda_0 = \Delta \cdot i \ll 1$ is valid, the difference $S_i(L_0) - S_i(L_0 + \lambda_0/2n)$ can be expressed as $2\pi\Delta \cdot i/\lambda_0 \cdot \sin(4\pi nL/\lambda_i + \gamma)$. Comparing it with the second term in (5) it's clear that for

$$\delta \approx \lambda_0/2nL_0 \quad (6)$$

the registered spectrum $S'_i$ will be erroneously approximated by the expression (2) with parameter $L_R = L_0 + \lambda_0/2n$, resulting in an abrupt error. However, in practical systems the wavelength scanning speed deviations are much weaker.

Therefore, the influence of the scale factor can be omitted for the considered interrogation method. In section 2.3 results of a numeric simulation, supporting this statement, are briefly discussed.

Investigating the third mechanism, let us consider the wavelength variation during the spectrum acquisition, resulting in $S'_i = S(L_0, \lambda_i + \delta\lambda)$, detailed as

$$S'_i \cong S_i(L_0) - S_m 4\pi nL_0 \delta\lambda_i / \lambda_0^2 \sin(4\pi nL_0/\lambda_i + \gamma). \quad (7)$$

As was mentioned, for all data processing we have utilized approach [8], in which only the slopes of the $S'_i$ spectrum are analyzed. It can be easily shown that at these spectral intervals $\sin(4\pi nL_0/\lambda_i + \gamma)$ function is close to unity. This will result in transformation of (7) to a form

$$S'_i \cong S_i(L_0) - S_m 4\pi nL_0 \delta\lambda_i / \lambda_0^2 . \quad (8)$$

Therefore, the value of SNR produced by the wavelength jitter will be expressed as

$$\text{SNR}_3 = \frac{S_m^2/2}{\left(\sigma_{\delta\lambda} S_m 4\pi nL_0 / \lambda_0^2\right)^2} = \frac{2\lambda_0^4}{\left(8\pi nL_0 \sigma_{\delta\lambda}\right)^2}, \quad (9)$$

where $\sigma_x$ denotes the standard deviation of $x_i$ array.

For consideration of the forth mechanism the measured interferometer spectral function $S'_i$ can be expressed as

$$S'_i \cong 2\left(R_1 R_2^*\right)^{1/2} \cos(4\pi nL_0/\lambda_i + \gamma)\cdot P_0 + \delta s_i, \quad (10)$$

where $P_0$ is the light source power. For simplicity an assumption of a Gaussian profile was applied to the fiber mode and to the light beam propagating inside the cavity. For this case the resulting expression for $R_2^* = R_2 \cdot \eta$ with $\eta$ as derived in Appendix A can be written as

$$R_2^*(\lambda, L) = R_2 \frac{(\pi n w_0^2)^2}{L^2\lambda^2 + (\pi n w_0^2)^2}. \quad (11)$$

In general, the standard deviation of the additive noises $\delta s_i$ can depend on the mean optical power incident to the photodetector. A simple yet practical approximation by a power function can be applied

$$\sigma_s = aP^b, \quad (12)$$

where $P = P_0 \cdot (R_1 + R_2^*)$ is the mean optical power incoming to the photodetector. The parameters $a$ and $b$ must be obtained explicitly for a given experimental setup. On this basis, the final expression for signal to noise ratio stipulated by the 4-th mechanism can be expressed as follows

$$\mathrm{SNR}_4 = \frac{S_m^2/2}{\sigma_s^2} = \frac{2P_0^{2-2b}}{a^2}\frac{R_1 R_2^*}{(R_1+R_2^*)^{2b}}. \quad (13)$$

It should be noted that for some particular values of parameter $b$, the expression (13) can be simplified. The first specific case we would like to consider is noise level independent of the incident optical power ($b=0$). Then the noise equivalent power of the photodetector can be introduced as $\mathrm{NEP} = \sigma_s$, appearing as a replacement for the (12). Expression for the $\mathrm{SNR}_4$ can therefore be written as

$$\mathrm{SNR}_4 = 2D^2 R_1 R_2^*(L), \quad (14)$$

where $D = P_0/\mathrm{NEP}$ is a dynamic range of the interrogating unit.

If the additive noise level is linearly related with the optical power (which is the case when the intensity noise of the light source is dominating), it's convenient to introduce RIN of the optical source (integrated over the whole frequency band of the photodetector). Then (12) is modified to $\sigma_s = \mathrm{RIN} \cdot P$, resulting in a following expression for $\mathrm{SNR}_4$

$$\mathrm{SNR}_4 = \frac{2R_1 R_2^*}{\mathrm{RIN}^2 \cdot (R_1+R_2^*)^2} = 0.5 \cdot \left(\frac{V}{\mathrm{RIN}}\right)^2, \quad (15)$$

where $V = S_m/P$ is the visibility of the EFPI spectrum fringes.

For all cases of $\mathrm{SNR}_4$ the resultant SNR can be calculated as

$$\mathrm{SNR} = 1/(\mathrm{SNR}_3^{-1} + \mathrm{SNR}_4^{-1}). \quad (16)$$

### 2.3. Investigation of the signal processing stability by means of numeric simulations

This section is devoted to study several features of the signal processing approach [8]; that is, to test the stability with additive noises and support Eqs. (4) and (6) concerning the influence of the wavelength scale distortions on the approximation results.

Estimation of the SNR is an intermediate step for determining the standard deviation of $L_R$ obtained by least-squares approximation. Therefore, the next step is to determine the noise-immunity of the utilized EFPI baseline calculation approach. For that purpose a numeric simulation with the following parameters was carried out: wavelength scanning range [1510; 1590] nm, $\Delta = 4$ pm, resulting in $M = 20001$ points in spectrum. A set of EFPI spectra $S'_i$ for cavity lengths from 30 μm to 1 mm was calculated according to expression (10). Additive noises $\delta s_i$ were simulated as an array of the same size as $\lambda_i$ (20001 points for our case) of normally distributed uncorrelated random quantities. Standard deviations of array $\delta s_i$ were set so that SNR of simulated spectra were varying in the range from 50 to 10000 (~17 to 40 dB). For each combination of the cavity length $L_0$ and SNR 1000 realizations of $S'_{ik}$ ($k = 1...1000$ – realization number) and a set of corresponding cavity length values $L_{Rk}$ were calculated. In figure 1 the relation of $\sigma_{Lr} = \mathrm{stdev}\{L_{Rk}\}$ and SNR value is illustrated. For $L_0 > 30$ μm the dependency of $\sigma_L(L_0)$ was quite weak, so an average over cavity lengths in the range [30, 1000] μm is shown.

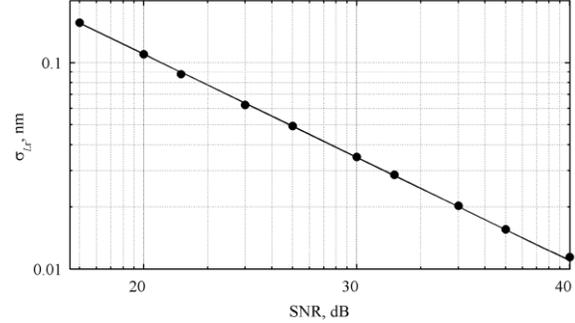

Figure 1. Relation of $L_R$ stdev and SNR of additive white noise.

The resultant dependency $\sigma_{Lr}(\mathrm{SNR})$ was approximated by power function

$$\sigma_{Lr}(\mathrm{SNR}) = C \cdot \mathrm{SNR}^{-1/2}, \quad (17)$$

$C = 1.1 \cdot 10^{-3}$ [μm] is conditioned by fitting approach [8] and can vary for other methods. The structure of the expression (17) is quite general, with square root dependency on the SNR, introduced for powers of signal and noise; constant $C$ dependent on the system parameters (number of spectral points $M$, signal processing approach). The obtained value $C$ is close to the Cramer-Rao bound for estimating the argument of a noisy sinusoid ([20], expression (23)), which for $M = 20001$ and recalculation of the argument deviation to the $L_R$ deviation gives a value $C \approx 9 \cdot 10^{-4}$ [μm]. The difference is likely to be caused by not exact correspondence of the EFPI baseline measurement problem to the problem of estimating the phase of a noisy sinusoid.

This result is necessary for estimation of the final $L_R$ fluctuations stipulated by non-ideal operation of the optical spectrum analyzer. Joining the effects of the scale shift (4) and "noisy factors" by summing corresponding variances and taking into account (17), the resultant standard deviation of the $L_R$ can be expressed as

$$\sigma_{Lr} = \left[\frac{C^2}{\mathrm{SNR}} + \frac{\sigma_{\Delta\lambda}^2 L_0^2}{\lambda_0^2}\right]^{1/2}. \quad (18)$$

The final expression for $\sigma_{Lr}$ can easily be obtained according to (18) with the use of (9), (11) and (13) for SNR estimation in (16), however, due to the bulkiness of the resulting formulae, it isn't presented in even form.

The investigation of the scale shift and the scale factor mechanisms was performed in a similar way. The parameters of the simulation were the same as the above-mentioned. For the scale shift influence examination, the spectra were calculated for a shifted wavelength ranges $\lambda + \Delta\lambda_0$, while in the approximation procedure initial wavelength range was substituted ($\Delta\lambda_0$ was of a practical values 0.01-1 pm). The dependency of the error of the obtained baseline value $\delta L(L)$ was in a perfect agreement with expression (4). In order to support the neglect of the scale factor influence, another test was performed – the spectra were calculated for a slightly extended spectral range with the same central point $\lambda_0$ but of a length $\Lambda+\Delta\lambda$, $\Delta\lambda$ was 0.01-1 pm, corresponding to $\delta \sim 10^{-7}\text{-}10^{-5}$. Again, the initial spectral range was substituted to the approximation procedure. The resulting dependency $\delta L(L)$ was approximated as $\delta L \approx 6.4\cdot 10^{-4}\cdot\delta\cdot L$. As a result, comparing the scale shift and scale factor influences, for equal values of the shift $\Delta\lambda_0$ and overall spectral range elongation $\Delta\lambda$, the baseline error produced by the scale factor is more than two orders less than the one produced by the scale shift. Therefore, the scale factor influence can be neglected without the loss of correctness.

## 3. Experimental study

In order to support the theoretical results, an experimental study of EFPI displacement sensor resolution was carried out. Spectra measurements were performed using the optical sensor interrogator NI PXIe 4844, utilizing a tunable laser with SMF-28 single-mode fiber output. Spectrometer parameters are the following: scanning range [1510; 1590] nm; $\Delta = 4$ pm; wavelength jitter stdev $\sigma_{\delta\lambda} = 1$ pm; optical power $P_0 \approx 0.06$ mW; scale shift stdev $\sigma_{\Delta\lambda} \approx 0.05$ pm.

In order to provide the relation between the additive noises and the incident optical power, a distinct experiment was performed – the level of optical power, reflected back to the interrogator was controlled by adjusting the mirrors with different reflectivities to the fiber end. This was done in order to test the system light source-photodetector itself, avoiding the interference, and hence, the presence of, for example, the distortions of the measured spectrum induced by the wavelength jitter. The resulting relation of the additive noise level $\sigma_s$ and the mean optical power $P$ was approximated by power function (12), where the fitted parameters are $a \approx 8.47\cdot 10^{-4}$ and $b \approx 0.81$. The experimental and the fitted results are shown in figure 2.

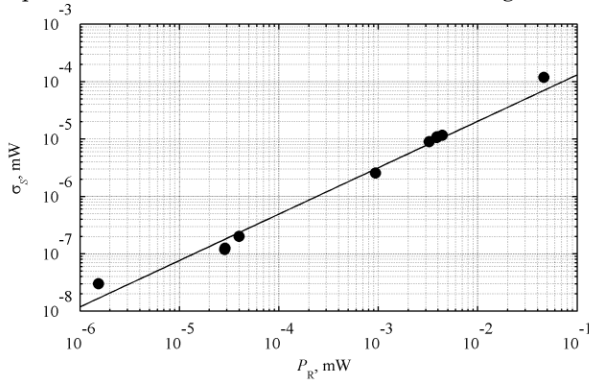

Figure 2. Relation of the optical power, reflected to the interrogator and the additive noise level – measured (points) and fitted (line).

Examined interferometer was formed by the two ends of SMF-28 fiber packaged with PC connectors, fixed in a standard mating sleeve. The air gap $L_0$ between the fiber ends was varied from ~30 μm up to 5 mm by the use of Standa 7TF2 translation stage. The radius of the used sleeve was 1.25 mm, greater than the effective radius of a Gaussian beam, travelled even several mm from the radiating fiber ($w(L)|_{L=10\text{ mm}} \approx$ 950 μm according to (A2)). This ensured the validity of free-space propagation approximation for the current experiment. The experimental setup is schematically illustrated in figure 3.

The left fiber was rigidly screwed to the sleeve, while the right one was fixed by the friction in the tube. The right fiber was connected to the translation stage only for the moments of $L_0$ adjustments and then was unhooked in order to eliminate the influences of the stage's possible mechanical vibrations. Three combinations of the fiber ends reflectivities were used: both 3.5% (Fresnel reflections at the glass-air boundary); 3.5% and ~20% (an increased reflectivity was produced by the dielectric evaporation on the fibre end); and 3.5% and ~90% (opaque aluminum mirror, glued to the fiber end). The interferometer was placed in a thermally isolated chamber in order to estimate the intrinsic limits of the measurement resolution. Also for this purpose the far end of the right fiber was put in the index-matching gel to avoid parasitic reflections.

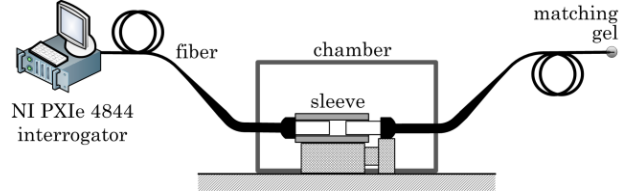

Figure 3. Experimental setup.

Spectra measurements were performed for about 10 minutes for each $L_0$ value, resulting in 600 spectra per $L_0$ point. For each measured spectrum it's fringe visibility and signal to noise ratio were calculated, averaged over all realizations and compared with the analytical predictions made according to (A8) and (16), with the following parameters substituted: $w_0 = 5.2$ μm, $\lambda_0 = 1.55$ μm, $n = 1$ and the values of $a$ and $b$ parameters for the expression (12) as indicated above. The results of this comparison are presented in figure 4. A good correspondence between the experimental and analytical results proves the adequacy of the developed model for coupling coefficient η, presented in Appendix A.

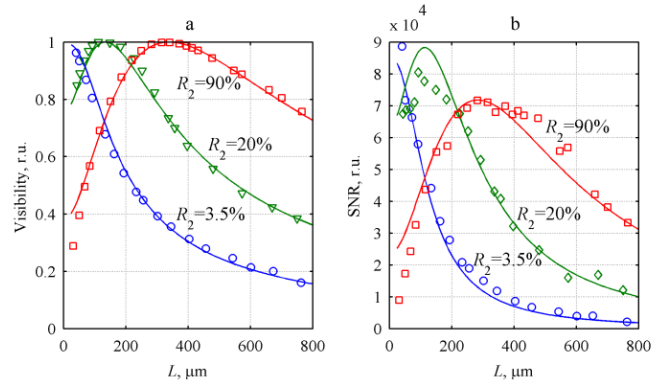

Figure 4. Fringe visibility (a) and signal to noise ratio (b) for different reflectivity of the second mirror; experimental (points) and analytical, calculated according to (A8) and (16) (solid curves).

The final step of the experimental study was to investigate the resolution of the baseline measurements. $L_{Rk}$ values were estimated with the use of approach [8] for each acquired spectrum $S'_{ik}$, $k = 1\ldots 600$. The resulting standard deviations $\sigma_{Lr}$ are shown in figure 5 by points, analytical predictions according to (18) are shown by curves.

The parameters substituted to (9), (11) and (13), and then to (18) were the following: $\lambda_0=1.55$ μm, $w_0=5.2$ μm (mode field radius for SMF-28 fiber at $\lambda_0=1.55$ μm), $\sigma_{\Delta\lambda}=0.05$ pm, $\sigma_{\delta\lambda}=1$ pm, $n = 1$, $a=8.47\cdot 10^{-4}$ and $b=0.81$ in (13), and reflections corresponding to the experimental ($R_1$=0.035, $R_2$=[0.035, 0.2, 0.9]).

Comparing the influence of different factors (scale shift, wavelength jitter, laser intensity and photodetector noises), one can conclude that in a given setup the influences of the scale shift and all noises are compatible for relatively large baselines. For instance, for $L$=500 µm, $R_1$=$R_2$=3.5%, the scale shift $\sigma_{\Delta\lambda}$=0.05 pm results in $\sigma_{L}$≈16 pm, while the overall noise influence produces $\sigma_{L}$≈17 pm, with resulting $\sigma_{L}$≈23 pm, which is in a good agreement with the experiment.

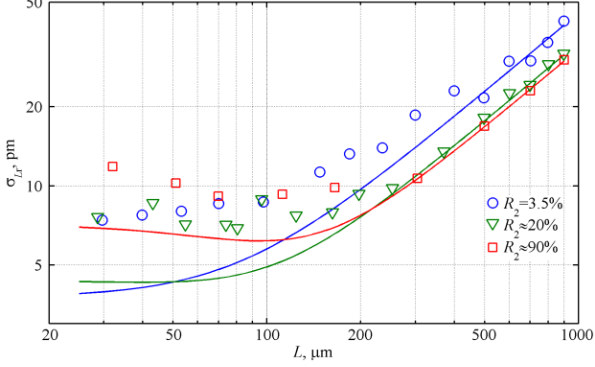

Figure 5. Baseline measurement resolutions for different reflectivity of the second mirror; experimental (points) and theoretical, calculated according to (18) (solid curves).

Estimating the resolution of a displacement sensor as $2\sigma_{Lr}$, the best resolution achieved in the described experiment was 14 pm (for cavity length 80 µm and $R_2 \approx 20\%$). For the $R_1 \approx 3.5\%$ the best attained standard deviation was ~7.5 pm, corresponding to 15 pm baseline resolution, at $L_0 \sim 30$ µm. For the $R_2 \approx 90\%$ the best attained standard deviation was ~9 pm, corresponding to 18 pm baseline resolution, at $L_0 \sim 70$-100 µm. The general behavior of experimental and analytical $\sigma_{Lr}(L)$ dependencies are in a very good agreement, however, for the baseline range, for which the expected standard deviations were 4-6 pm, the attained stdevs were 8-10 pm instead.

### 4. Discussion and conclusion

The developed model is applicable for estimation of the EFPI displacement sensors resolution limits induced by the acquisition hardware. However, for calculation of parameters of a particular experimental setup, some specific characteristics of interrogating unit must be specified. For example, in the current study a distinct experiment for estimating the parameters, relating the mean optical power and the additive noises $a$ and $b$ was performed.

In the performed experiment a picometer-level resolution was attained. The experimental results are in a good correspondence with the analytical expectations. Such secondary characteristics as signal to noise ratio and fringe visibility of the measured EFPI spectra are in an extremely good agreement with the expectations given by our model. The baseline resolution is also in a good agreement with theory in the range $L > 100$ µm. However, at the cavity length range from ~30 µm to ~100 µm the achieved baseline standard deviation was ~8-10 pm instead of ~4-7 pm, yet the main behavior was quite similar. Such excess of the experimental values over the analytical predictions can be explained by additional uncorrelated fluctuations of the $L_0$ with standard deviation about 7 pm for all the three assembled EFPI configurations used in our experiments. This is most likely to be caused by the intrinsic $L_0$ deviations of thermo-mechanical nature. However, even despite this slight disagreement of experiment and analytical predictions, the developed model provides a good description of EFPI sensors resolution even for short cavities, which are the main trend in the recent works [4-6,14].

The developed model can be applied to other types of interrogating setups, such as spectrometers based on the step-tuned lasers and utilizing broad-band source and tunable high-finesse optical filter.

Coincidence of theoretical and experimental results proves the adequacy of the developed model for the purpose of analytical estimation of the possible EFPI displacement sensor resolution, enabling one to estimate the possible resolution limit for a given setup, or to derive the requirements for optical elements and/or interrogator necessary for attaining a desired sensor resolution. Also a baseline resolution of 14-15 pm was attained.

### Appendix A

One of the key problems for the desired model is description of the light beam passed through a low-finesse Fabry-Perot cavity, coupled to a single-mode fiber. In this case the light propagation (forward, reflection and backward propagation) can be considered as travelling the distance equal to doubled cavity length $z = 2L$. For both the fiber mode and the optical beam propagating inside the cavity an approximation of the Gaussian profile was applied, for which the distribution of an electric field complex amplitude can be written as [21]

$$A(r,z) = \frac{w_0}{w(z)} \exp\left[-\frac{r}{w(z)} - j\frac{z \cdot r^2}{z_R w^2(z)} - j\operatorname{atan}\left(\frac{z}{z_R}\right)\right], \quad (A1)$$

where beam radius $w(z)$ and the so-called Rayleigh length $z_R$ are given by expressions

$$w(z) = w_0\sqrt{1 + (z/z_R)^2}, \quad z_R = \frac{\pi n w_0^2}{\lambda}. \quad (A2)$$

Herein, the distribution of the optical beam reflected from the interferometer and incident to the fiber can be approximated as $A(r, 2L)$, and the fiber mode profile is expressed as $A(r, 0)$. This approximation may be not quite accurate for the case of large $L$, when the beam radius becomes greater than the radius of the mirror (in our case the main reflector is fiber core and cladding with diameter of 125 µm), and therefore, the Gaussian beam geometry is distorted. However, this distortion will happen to the boundary parts of the beam, while in our task the principal part is located near the beam axis (closer than ~5 µm according to the fiber MFD). Provided that, the effect of the finite mirror on the field in the central part of the beam is negligible, and thus using this assumption gives an accurate result.

In order to take into account the diffraction-induced optical losses, an overlapping coefficient of the fiber mode and the optical beam, incident to the first fiber must be calculated. For the electric field this coefficient is given by expression

$$\eta_F(L) = \frac{\iint A(r,0) \cdot A^*(r,2L) d\varphi dr}{\left[\iint |A(r,0)|^2 d\varphi dr \iint |A(r,2L)|^2 d\varphi dr\right]^{1/2}}. \quad (A3)$$

Substituting (A1) into (A3) and evaluating the integrals, one obtains the electric fields overlapping coefficient in the following form

$$\eta_F = \frac{2w_0 w(2L) \cdot e^{-j\psi}}{\left[\left(w_0^2 + w^2(2L)\right)^2 + 4w_0^4 L^2/z_R^2\right]^{1/2}}, \quad (A4)$$

where the phase term $\psi$ is

$$\psi = \operatorname{atan}\left[\frac{L}{z_R} \cdot \left(4\frac{L^2}{z_R^2} + 3\right)\right]. \quad (A5)$$

The coupling coefficient for the optical power can be calculated as $\eta = |\eta_F|^2$, resulting in a following expression

$$\eta = \frac{\left(\pi n w_0^2\right)^2}{L^2\lambda^2 + \left(\pi n w_0^2\right)^2}. \quad (A6)$$

An expression of similar sense for optical power coupling was obtained in [22], however, the presented form is more convenient for consideration of the EFPI.

The amplitude $S_m$ of the oscillating part of the EFPI spectral function (2) can be written as

$$S_m = \frac{\pi n w_0^2 \sqrt{R_1 R_2}}{\sqrt{L^2\lambda^2 + \left(\pi n w_0^2\right)^2}}. \quad (A7)$$

When considering the whole spectrum (including constant component), such characteristic as fringe visibility may be useful. Taking into account (A6), it can be expressed as

$$V = \frac{2\sqrt{R_1 R_2}}{R_1 \cdot \left[\left(L\lambda/\pi n w_0^2\right)^2 + 1\right]^{1/2} + R_2}. \quad (A8)$$

The parameters $V$ and $S_m$ can relatively easily be determined from a measured spectral function of the interferometer and their dependencies on the baseline $V(L)$ and $S_m(L)$ can therefore be used to prove the adequacy of the developed model.

## Appendix B

The phase term $\gamma(L)$, added to the geometrical phase delay, is considered in this Appendix. This term is given by the expression $\gamma(L) = -\psi(L) + \varphi$, $\psi$ is given by (A5) and $\varphi$ is a phase shift, induced by the mirrors. For the case of dielectric mirrors a simplification $\varphi = \pi$ can be made.

The term $\psi(L)$ mostly affect the absolute value of the resulting OPD, while its influence on the resolution of the EFPI sensor is negligible. For instance, the influence of the wavelength jitter mechanism can be investigated by comparing the derivatives of all the terms under the cosine function with respect to $\lambda$:

$$\left.\frac{\partial}{\partial\lambda}\frac{4\pi n L}{\lambda}\right|_{\lambda_0} = \frac{4\pi n L}{\lambda_0^2}, \quad (A9)$$

$$\left.\frac{\partial}{\partial\lambda}\psi(L)\right|_{\lambda_0} = \frac{3\left(\pi n w_0^2\right)^3 \cdot L}{\left[(L\lambda)^2 + \left(\pi n w_0^2\right)^2\right] \cdot \left[(2L\lambda)^2 + \left(\pi n w_0^2\right)^2\right]}. \quad (A10)$$

The calculations of the above expressions were made with the following parameters: $w_0 = 5.2\ \mu m$, $\lambda_0 = 1.55\ \mu m$, $n = 1$, $L$ from 20 μm to 100 μm. According to such calculations one can observe that the (A9) is two-three orders greater than the (A10). As can be seen from the asymptotics of the (A10), the $\psi$ term becomes nearly constant with respect to wavelength for relatively large $L > 50\ \mu m$. Therefore, in the consideration of the $\Delta\lambda_0$ and $\delta\lambda_i$ noisy mechanisms the $\psi(L)$ term can be omitted without the loss of correctness.